\documentclass[twocolumn,showpacs,amsmath,amssymb,superscriptaddress]{revtex4}
\usepackage{graphicx}
\usepackage{dcolumn}
\usepackage{bm}
\begin{document}

\title{Error Correcting Codes For Adiabatic Quantum Computation}
\author{Stephen P. Jordan}
\email{sjordan@mit.edu}
\author{Edward Farhi}
\affiliation{Center for Theoretical Physics, Massachusetts Institute
  of Technology, Cambridge, Massachusetts 02139}
\author{Peter W. Shor}
\affiliation{Mathematics Department, Massachusetts Institute of
  Technology, Cambridge, Massachusetts 02139}
\date{\today}

\newcommand{\ud}{\mathrm{d}}
\newcommand{\bra}[1]{\langle #1|}
\newcommand{\ket}[1]{|#1\rangle}
\newcommand{\braket}[2]{\langle #1|#2\rangle}

\begin{abstract}
Recently, there has been growing interest in using adiabatic quantum
computation as an architecture for experimentally realizable quantum
computers. One of the reasons for this is the idea that the energy gap
should provide some inherent resistance to noise. It is now known that
universal quantum computation can be achieved adiabatically using
2-local Hamiltonians. The energy gap in these Hamiltonians scales as
an inverse polynomial in the problem size. Here we present stabilizer
codes which can be used to produce a constant energy gap against
1-local and 2-local noise. The corresponding fault-tolerant universal
Hamiltonians are 4-local and 6-local respectively, which is the
optimal result achievable within this framework.
\end{abstract}

\pacs{03.67.Pp}
\maketitle

Adiabatic quantum computation was originally proposed by Farhi
\emph{et al.} as a method for solving combinatorial optimization
problems \cite{Farhi}. In this scheme, one starts with a Hamiltonian
whose ground state is easy to construct, and gradually varies the
Hamiltonian into one whose ground state encodes the solution to a
computational problem. By the adiabatic theorem, the system will
remain in the instantaneous ground state provided that the Hamiltonian
is varied sufficiently slowly. More precisely, any closed system acted
on by $H(t/T)$ from $t=0$ to $T$ will remain in the ground state with
high probability provided that $T$ is sufficiently large. Different
formulations\cite{Messiah, Jansen, Schaller, Joye}  of the adiabatic
theorem yield different conditions on $T$, but essentially the minimal
$T$ scales polynomially with the inverse eigenvalue gap between the
ground state and first excited state.

Recently, there has been growing interest in using adiabatic quantum
computation as an architecture for experimentally realizable quantum
computers. Aharonov \emph{et al.}\cite{Aharonov}, building on ideas by
Feynman\cite{Feynman} and Kitaev\cite{Kitaev}, showed that any quantum
circuit can be simulated by an adiabatic quantum algorithm. The energy
gap for this algorithm scales as an inverse polynomial in $G$,  the
number of gates in the original quantum circuit. $G$ is identified as
the running time of the original circuit. By the adiabatic theorem,
the running time of the adiabatic simulation is polynomial in
$G$. Because the slowdown is only polynomial, adiabatic quantum
computation is a form of universal quantum computation.

Most experimentally realizable Hamiltonians involve only few-body
interactions. Thus theoretical models of quantum computation are
usually restricted to involve interactions between at most some
constant number of qubits $k$. Any Hamiltonian on $n$ qubits can be
expressed as a linear combination of terms, each of which is a tensor
product of $n$ Pauli matrices, where we include the $2 \times 2$
identity as a fourth Pauli matrix. If each of these tensor products
contains at most $k$ Pauli matrices not equal to the identity then the
Hamiltonian is said to be $k$-local. The Hamiltonian used in the
universality construction of \cite{Aharonov} is 3-local throughout the
time evolution. Kempe \emph{et al.} subsequently improved this to
2-local in \cite{Kempe}.

Schr\"odinger's equation shows that, for any constant $g$, $g H(g t)$
yields the same time evolution from time $0$ to $T/g$ that $H(t)$
yields from $0$ to $T$. Thus, the running time of an adiabatic
algorithm would not appear to be well defined. However, in any
experimental realization there will be a limit to the magnitude of the
fields and couplings. Thus it is reasonable to limit the norm of each
term in $H(t)$. Such a restriction enables one to make statements
about how the running time of an adiabatic algorithm scales with some
measure of the problem size, such as $G$. 

One of the reasons for interest in adiabatic quantum computation
as an architecture is the idea that adiabatic quantum computers may
have some inherent fault tolerance \cite{Childs, Sarandy, Aberg,
  Roland, Kaminsky} . Because the final state depends only on the final
Hamiltonian, adiabatic quantum computation may be resistant to slowly
varying control errors, which cause $H(t)$ to vary from its intended
path, as long as the final Hamiltonian is correct. An exception to this
would occur if the modified path has an energy gap small enough to
violate the adiabatic condition. Unfortunately, it is generally quite
difficult to evaluate the energy gap of arbitrary local Hamiltonians.

Another reason to expect that adiabatic quantum computations may be
inherently fault tolerant is that the energy gap should provide some
inherent resistance to noise caused by stray couplings to the
environment. Intuitively, the system will be unlikely to get excited
out of its ground state if $k_b T$ is less than the energy
gap. Unfortunately, in most proposed applications of adiabatic quantum
computation, the energy gap scales as an inverse polynomial in the
problem size. Such a gap only affords protection if the temperature
scales the same way. However, a temperature which shrinks polynomially
with the problem size may be hard to achieve experimentally.

To address this problem, we propose taking advantage of the
possibility that the decoherence will act independently on the
qubits. The rate of decoherence should thus depend on the energy gap
against local noise. We construct a class of stabilizer codes such
that encoded Hamiltonians are guaranteed to have a constant energy gap
against single-qubit excitations. These stabilizer codes are designed
so that adiabatic quantum computation with 4-local Hamiltonians is
universal for the encoded states. We illustrate the usefulness of
these codes for reducing decoherence using a noise model, proposed in
\cite{Childs}, in which each qubit independently couples to a photon
bath.

To protect against decoherence we wish to create an energy gap against
single-qubit disturbances. To do this we use a quantum error
correcting code such that applying a single Pauli operator to any
qubit in a codeword will send this state outside of the codespace. Then we
add an extra term to the Hamiltonian which gives an energy penalty to
all states outside the codespace. Since we are only interested in
creating an energy penalty for states outside the codespace, only the
fact that an error has occurred needs to be detectable. Since we are
not actively correcting errors, it is not necessary for distinct
errors to be distinguishable. In this sense, our code is not truly an
error correcting code but rather an error \emph{detecting} code. Such
passive error correction is similar in spirit to ideas suggested for
the circuit model in \cite{Bacon}.

It is straightforward to verify that the 4-qubit code
\begin{eqnarray}
\label{zero_logical}
\ket{0_L} & = & \frac{1}{2} \left( \ket{0000} + i\ket{0011}
+i\ket{1100} + \ket{1111} \right)
\\
\label{one_logical}
\ket{1_L} & = & \frac{1}{2} \left( -\ket{0101} + i\ket{0110} +
i\ket{1001} - \ket{1010} \right)
\end{eqnarray}
satisfies the error-detection requirements, namely
\begin{equation}
\label{detection}
\bra{0_L} \sigma \ket{0_L} = \bra{1_L} \sigma \ket{1_L} = \bra{0_L}
\sigma \ket{1_L} = 0
\end{equation}
where $\sigma$ is any of the three Pauli operators acting on one
qubit. Furthermore, the following 2-local operations act as encoded
Pauli $X$, $Y$, and $Z$ 
operators.
\begin{equation}
\label{logical_operators}
\begin{array}{lll}
X_L & = & Y \otimes I \otimes Y \otimes I
\\
Y_L & = & -I \otimes X \otimes X \otimes I
\\
Z_L & = & Z \otimes Z \otimes I \otimes I
\end{array}
\end{equation}
That is,
\[
\begin{array}{llllll}
X_L \ket{0_L} & = & \ket{1_L}, & X_L \ket{1_L}   & = & \ket{0_L}, \\
Y_L \ket{0_L} & = & i \ket{1_L}, & Y_L \ket{1_L} & = & - i \ket{0_L}, \\
Z_L \ket{0_L} & = & \ket{0_L}, & Z_L \ket{1_L}   & = & - \ket{1_L}.
\end{array}
\]
An arbitrary state of a single qubit $\alpha \ket{0} + \beta \ket{1}$
is encoded as $\alpha \ket{0_L} + \beta \ket{1_L}$. 

Starting with an arbitrary 2-local Hamiltonian $H$ on $N$ bits, we
obtain a new fault tolerant Hamiltonian on $4N$ bits by the following
procedure. An arbitrary 2-local Hamiltonian can be written as a sum of
tensor products of pairs of Pauli matrices acting on different
qubits. After writing out $H$ in this way, make the following
replacements
\[
\begin{array}{llll}
I \to I^{\otimes 4}, & X \to X_L, & Y \to Y_L, & Z \to Z_L
\end{array}
\] 
to obtain a new 4-local Hamiltonian $H_{SL}$ acting on $4N$
qubits. The total fault tolerant Hamiltonian $H_S$ is
\begin{equation}
\label{total}
H_S = H_{SL} + H_{SP}
\end{equation}
where $H_{SP}$ is a sum of penalty terms, one acting on each encoded
qubit, providing an energy penalty of at least $E_p$ for going outside
the code space. We use the subscript $S$ to indicate that the
Hamiltonian acts on the system, as opposed to the environment, which
we introduce later. Note that $H_{SL}$ and $H_{SP}$ commute, and thus
they share a set of simultaneous eigenstates.

If the ground space of $H$ is spanned by  $\ket{\psi^{(1)}} \ldots
\ket{\psi^{(m)}}$ then the ground space of $H_S$ is spanned by the
encoded states $\ket{\psi_L^{(1)}} \ldots
\ket{\psi_L^{(m)}}$. Furthermore, the penalty terms provide an energy
gap against 1-local noise which does not shrink as the size of the
computation grows.

The code described by equations \ref{zero_logical} and
\ref{one_logical} can be obtained using the stabilizer formalism
\cite{Gottesman, Nielsen}. In this formalism, a quantum code is not
described by explicitly specifying a set of basis states for the code
space. Rather, one specifies the generators of the stabilizer group for
the codespace. Let $G_n$ be the Pauli group on $n$ qubits (\emph{i.e.}
the set of all tensor products of $n$ Pauli operators with 
coefficients of $\pm 1$ or $\pm i$). The stabilizer group of a
codespace $C$ is the subgroup $S$ of $G_n$ such that $x \ket{\psi} =
\ket{\psi}$ for any $x \in S$ and any $\ket{\psi} \in C$.

A $2^k$ dimensional codespace over $n$ bits can be specified by
choosing $n-k$ independent commuting generators for the stabilizer
group $S$. By independent we mean that no generator
can be expressed as a product of others. In our case we are encoding a
single qubit using 4 qubits, thus $k=1$ and $n=4$, and we need 3
independent commuting generators for $S$. 

To satisfy the orthogonality conditions, listed in equation
\ref{detection}, which are necessary for error detection, it suffices
for each Pauli operator on a given qubit to anticommute with at least
one of the generators of the stabilizer group. The generators
\begin{eqnarray}
\label{generators}
g_1 & = & X \otimes X \otimes X \otimes X \nonumber \\
g_2 & = & Z \otimes Z \otimes Z \otimes Z \nonumber \\
g_3 & = & X \otimes Y \otimes Z \otimes I
\end{eqnarray}
satisfy these conditions, and generate the stabilizer group for
the code given in equations \ref{zero_logical} and \ref{one_logical}.

Adding one term of the form
\begin{equation}
\label{penalty}
H_p = -E_p (g_1+g_2+g_3)
\end{equation}
to the encoded Hamiltonian for each encoded qubit yields an energy
penalty of at least $E_p$ for any state outside the codespace.

2-local encoded operations are optimal. None of the encoded
operations can be made 1-local, because they would then have the same
form as the errors we are trying to detect and penalize. Such an
operation would not commute with all of the generators. 

Intuitively, one expects that providing an energy gap against a Pauli
operator applied to any qubit protects against 1-local noise. We
illustrate this using a model of decoherence proposed in
\cite{Childs}. In this model, the quantum computer is a set of
spin-$1/2$ particles weakly coupled to a large photon bath. The
Hamiltonian for the combined system is
\[
H = H_S + H_E + \lambda V,
\]
where $H_S(t)$ is the adiabatic Hamiltonian that implements the
algorithm by acting only on the spins, $H_E$ is the Hamiltonian which
acts only on the photon bath, and $\lambda V$ is a weak coupling
between the spins and the photon bath. Specifically, $V$ is assumed to
take the form
\[
V = \sum_i \int_0^\infty \ud \omega \left[ g(\omega) a_\omega
  \sigma_+^{(i)} + g^*(\omega)a_\omega^\dag \sigma_-^{(i)} \right],
\]
where $\sigma_{\pm}^{(i)}$ are raising and lowering operators for the
$i$th spin, $a_\omega$ is the annihilation operator for the photon
mode with frequency $\omega$, and $g(\omega)$ is the spectral density. 

From this premise Childs \emph{et al.} obtain the following master
equation 
\begin{equation}
\label{master_equation}
\frac{\ud \rho}{\ud t} = -i[H_S,\rho]-\sum_{a,b} M_{ab} \ 
\mathcal{E}_{ab}(\rho)
\end{equation}
where
\begin{eqnarray*}
M_{ab} & = &  \sum_i \left[ N_{ba}
  |g_{ba}|^2 \bra{a} \sigma_-^{(i)} \ket{b} \bra{b} \sigma_+^{(i)}
  \ket{a} \right. \\
  & & \left. + (N_{ab}+1) |g_{ab}|^2 \bra{b} \sigma_-^{(i)} \ket{a}
      \bra{a} \sigma_+^{(i)} \ket{b} \right]
\end{eqnarray*}
is a scalar,
\[
\mathcal{E}_{ab}(\rho) = \ket{a} \bra{a} \rho
+ \rho \ket{a} \bra{a} - 2 \ket{b} \bra{a} \rho \ket{a} \bra{b}
\]
is an operator, $\ket{a}$ is the instantaneous eigenstate of $H_S$
with energy $\omega_a$,
\[
N_{ba} = \frac{1}{\exp \left[ \beta (\omega_b-\omega_a)\right]-1}
\]
is the Bose-Einstein distribution at temperature $1/\beta$, and
\begin{equation}
\label{g}
g_{ba} = \left\{ \begin{array}{ll}
\lambda g(\omega_b - \omega_a), & \omega_b > \omega_a, \\
0, & \omega_b \leq \omega_a. \end{array} \right.
\end{equation}

Suppose that we encode the original $N$-qubit Hamiltonian as a
$4N$-qubit Hamiltonian as described above. As stated in equation
\ref{total}, the total spin Hamiltonian $H_S$ on $4N$ spins
consists of the encoded version $H_{SL}$ of the original Hamiltonian
$H_S$ plus the penalty terms $H_{SP}$.  

Most adiabatic quantum computations use an initial Hamiltonian with an
eigenvalue gap of order unity, independent of problem size. In such
cases, a nearly pure initial state can be achieved at constant
temperature. Therefore, we'll make the approximation that the spins
start in the pure ground state of the initial Hamiltonian, which we'll
denote $\ket{0}$. Then we can use equation \ref{master_equation} to
examine $\ud \rho/ \ud t$ at $t=0$. Since the initial state is $\rho =
\ket{0} \bra{0}$, $\mathcal{E}_{ab}(\rho)$ is zero unless $\ket{a} =
\ket{0}$. The master equation at $t=0$ is therefore 
\begin{equation}
\label{intermediate}
\left. \frac{\ud \rho}{\ud t}\right|_{t=0} = -i[H_S,\rho]-\sum_b
M_{0b} \ \mathcal{E}_{0b}(\rho).
\end{equation}

$H_{SP}$ is given by a sum of terms of the form \ref{penalty}, and it
commutes with $H_{SL}$. Thus, $H_S$ and $H_{SP}$ share a complete set
of simultaneous eigenstates. The eigenstates of $H_S$ can thus be
separated into those which are in the codespace $C$ (\emph{i.e.} the
ground space of $H_{SP}$) and those which are in the orthogonal space
$C^\perp$. The ground state $\ket{0}$ is in the codespace. $M_{0b}$
will be zero unless $\ket{b} \in C^\perp$, because $\sigma_{\pm} = (X
\pm i Y)/2$, and any Pauli operator applied to a single bit takes us
from $C$ to $C^\perp$. Equation \ref{intermediate} therefore becomes  
\begin{equation}
\label{master_equation2}
\left. \frac{\ud \rho}{\ud t} \right|_{t=0} = -i
     [H_S,\rho]+\sum_{b \in C^\perp} M_{0b} \ \mathcal{E}_{0b}(\rho)
\end{equation}

Since $\ket{0}$ is the ground state, $\omega_b \geq \omega_0$, thus
equation \ref{g} shows that the terms in $M_{0b}$ proportional to
$|g_{0b}|^2$ will vanish, leaving only
\[
M_{0b} = \sum_i N_{b0} |g_{b0}|^2 \bra{0}\sigma_-^{(i)}\ket{b} \bra{b}
\sigma_+^{(i)} \ket{0}.
\]

Now let's examine $N_{b0}$.
\[
\omega_b - \omega_0 = \bra{b} (H_{SL} + H_{SP}) \ket{b} - \bra{0}
(H_{SL} + H_{SP}) \ket{0}.
\] 
$\ket{0}$ is in the ground space of $H_{SL}$, thus 
\[
\bra{b} H_{SL} \ket{b} - \bra{0} H_{SL} \ket{0} \geq 0,
\]
and so
\[
\omega_b - \omega_0 \geq \bra{b} 
H_{SP} \ket{b} - \bra{0} H_{SP} \ket{0}.
\]
Since $\ket{b} \in C^\perp$ and $\ket{0} \in C$, 
\[
\bra{b} H_{SP} \ket{b} - \bra{0} H_{SP} \ket{0}
= E_p,
\]
thus $\omega_b - \omega_0 \geq E_p$.

A sufficiently large $\beta E_p$ will make $N_{ba}$ small enough that
the term $\sum_{b \in C^\perp} M_{0b} \mathcal{E}(\rho)$ can be
neglected from the master equation, leaving
\[
\left. \frac{\ud \rho}{\ud t} \right|_{t=0} \approx - i [H_S,\rho]
\]
which is just Schr\"odinger's equation with a Hamiltonian equal to
$H_S$ and no decoherence. Note that the preceding derivation did
not depend on the fact that $\sigma_\pm^{(i)}$ are raising and
lowering operators, but only on the fact that they act on a single
qubit and can therefore be expressed as a linear combination of Pauli
operators.

$N_{b0}$ is small but nonzero. Thus, after a sufficiently long time,
the matrix elements of $\rho$ involving states other than $\ket{0}$
will become non-negligible and the preceding picture will break down. How
long the computation can be run before this happens depends on the
magnitude of $\sum_{b \in C^\perp} M_{ob} \mathcal{E}(\rho)$, which
shrinks exponentially with $E_p/T$ and grows only polynomially
with the number of qubits $N$. Thus it should be sufficient for
$1/T$ to grow logarithmically with the problem size. In
contrast, one expects that if the Hamiltonian had only an inverse
polynomial gap against 1-local noise, the temperature would need to
shrink polynomially rather than logarithmically.

Now that we know how to obtain a constant gap against 1-local noise,
we may ask whether the same is possible for 2-local noise. To 
accomplish this we need to find a stabilizer group such that any pair
of Pauli operators on two bits anticommutes with at least one of the
generators. This is exactly the property satisfied by the
standard\cite{Nielsen} 5-qubit stabilizer code, whose stabilizer group
is generated by
\begin{eqnarray}
\label{fivegenerators}
g_1 & = & X \otimes Z \otimes Z \otimes X \otimes I \nonumber \\
g_2 & = & I \otimes X \otimes Z \otimes Z \otimes X \nonumber \\
g_3 & = & X \otimes I \otimes X \otimes Z \otimes Z \nonumber \\
g_4 & = & Z \otimes X \otimes I \otimes X \otimes Z. 
\end{eqnarray}
The codewords for this code are
\begin{eqnarray*}
\ket{0_L} & = & \frac{1}{4} \left[ \ \ket{00000} + \ket{10010} +
  \ket{01001} +\ket{10100} \right. \\
& & + \ket{01010} - \ket{11011} - \ket{00110} - \ket{11000} \\
& & - \ket{11101} -\ket{00011} - \ket{11110} - \ket{01111} \\
& & \left. - \ket{10001} - \ket{01100} - \ket{10111} + \ket{00101}
\  \right]
\end{eqnarray*}
\begin{eqnarray*}
\ket{1_L} & = & \frac{1}{4} \left[ \ \ket{11111} + \ket{01101} +
  \ket{10110} + \ket{01011} \right. \\
& & + \ket{10101} - \ket{00100} - \ket{11001} - \ket{00111} \\
& & - \ket{00010} - \ket{11100} - \ket{00001} - \ket{10000} \\
& & \left. - \ket{01110} - \ket{10011} - \ket{01000} + \ket{11010}
\ \right]. 
\end{eqnarray*}
The encoded Pauli operations for this code are conventionally
expressed as
\begin{eqnarray*}
X_L & = & X \otimes X \otimes X \otimes X \otimes X \\
Y_L & = & Y \otimes Y \otimes Y \otimes Y \otimes Y \\
Z_L & = & Z \otimes Z \otimes Z \otimes Z \otimes Z.
\end{eqnarray*}
However, multiplying these encoded operations by members of the
stabilizer group doesn't affect their action on the codespace. Thus we
obtain the following equivalent set of encoded operations.
\begin{eqnarray}
\label{five_logical}
X_L & = & -X \otimes I \otimes Y \otimes Y \otimes I  \nonumber \\
Y_L & = & -Z \otimes Z \otimes I \otimes Y \otimes I \nonumber \\
Z_L & = & -Y \otimes Z \otimes Y \otimes I \otimes I
\end{eqnarray}
These operators are all 3-local. This is the best that can be hoped
for, because the code protects against 2-local operations and
therefore any 2-local operation must anticommute with at least one of
the generators.

Besides increasing the locality of the encoded operations, one can
seek to decrease the number of qubits used to construct the
codewords. The quantum singleton bound\cite{Nielsen} shows that the
five qubit code is already optimal and cannot be improved in this
respect.

The distance $d$ of a quantum code is the minimum number of qubits of
a codeword which need to be modified before obtaining a nonzero inner
product with a different codeword. For example, applying $X_L$, which
is 3-local, to $\ket{0_L}$ of the 5-qubit code converts it into
$\ket{1_L}$, but applying any 2-local operator to any of the codewords
yields something outside the codespace. Thus the distance of the
5-qubit code is 3. Similarly the distance of our 4-qubit code is 2. To
detect $t$ errors a code needs a distance of $t+1$, and to correct $t$
errors, it needs a distance of $2t+1$.

The quantum singleton bound states that the distance of any quantum
code which uses $n$ qubits to encode $k$ qubits will satisfy
\begin{equation}
\label{singleton}
n-k \geq 2(d-1).
\end{equation}
To detect 2 errors, a code must have distance 3. A code which
encodes a single qubit with distance 3 must use at least 5 qubits, by
equation \ref{singleton}. Thus the 5-qubit code is optimal. To detect
1 error, a code must have distance 2. A code which encodes a single
qubit with distance 2 must have at least 3 qubits, by equation
\ref{singleton}. Thus it appears possible that our 4-qubit code is not
optimal. However, no 3-qubit stabilizer code can detect all
single-qubit errors, which we show as follows.

The stabilizer group for a 3-qubit code would have two independent
generators, each being a tensor product of 3 Pauli operators. 
\begin{eqnarray*}
g_1 & = & \sigma_{11} \otimes \sigma_{12} \otimes \sigma_{13} \\
g_2 & = & \sigma_{21} \otimes \sigma_{22} \otimes \sigma_{23}
\end{eqnarray*} \\
These must satisfy the following two conditions: (1) they commute, and
(2) an $X, Y$, or $Z$ on any of the three qubits anticommutes with at
least one of the generators. This is impossible, because condition (2)
requires $\sigma_{1i} \neq \sigma_{2i} \neq I$ for each
$i=1,2,3$. In this case $g_1$ and $g_2$ anticommute.

The stabilizer formalism describes most but not all currently known
quantum error correcting codes. We do not know whether a 3-qubit code
which detects all single-qubit errors while still maintaining 2-local
encoded operations can be found by going outside the stabilizer
formalism. It may also be interesting to investigate whether there
exist computationally universal 3-local or 2-local adiabatic
Hamiltonians with a constant energy gap against local noise.

We thank Ed Platt, Jay Gill, Shay Mozes, Daniel Lidar, and Mark Rudner
for useful discussions. We especially thank David DiVincenzo for
encouraging us to work on this topic, and Andrew Childs for helping to
clarify an important point. EF gratefully acknowledges support from
the National Security Agency (NSA) and Advanced Research and
Development Activity (ARDA) under Army Research Office (ARO) contract
W911NF-04-1-0216. SJ gratefully acknowledges support from ARO/ARDA's
QuaCGR program. PS gratefully acknowledges support from the NSF under
grant number CCF-0431787.

\bibliography{notes}

\begin{thebibliography}{17}
\expandafter\ifx\csname natexlab\endcsname\relax\def\natexlab#1{#1}\fi
\expandafter\ifx\csname bibnamefont\endcsname\relax
  \def\bibnamefont#1{#1}\fi
\expandafter\ifx\csname bibfnamefont\endcsname\relax
  \def\bibfnamefont#1{#1}\fi
\expandafter\ifx\csname citenamefont\endcsname\relax
  \def\citenamefont#1{#1}\fi
\expandafter\ifx\csname url\endcsname\relax
  \def\url#1{\texttt{#1}}\fi
\expandafter\ifx\csname urlprefix\endcsname\relax\def\urlprefix{URL }\fi
\providecommand{\bibinfo}[2]{#2}
\providecommand{\eprint}[2][]{\url{#2}}

\bibitem[{\citenamefont{Farhi et~al.}(2000)\citenamefont{Farhi, Goldstone,
  Gutmann, and Sipser}}]{Farhi}
\bibinfo{author}{\bibfnamefont{E.}~\bibnamefont{Farhi}},
  \bibinfo{author}{\bibfnamefont{J.}~\bibnamefont{Goldstone}},
  \bibinfo{author}{\bibfnamefont{S.}~\bibnamefont{Gutmann}}, \bibnamefont{and}
  \bibinfo{author}{\bibfnamefont{M.}~\bibnamefont{Sipser}},
  \bibinfo{journal}{arXiv:quant-ph/0001106}  (\bibinfo{year}{2000}).

\bibitem[{\citenamefont{Messiah}(1958)}]{Messiah}
\bibinfo{author}{\bibfnamefont{A.}~\bibnamefont{Messiah}},
  \emph{\bibinfo{title}{Quantum Mechanics}} (\bibinfo{publisher}{Dover},
  \bibinfo{year}{1958}).

\bibitem[{\citenamefont{Jansen et~al.}(2006)\citenamefont{Jansen, Ruskai, and
  Seiler}}]{Jansen}
\bibinfo{author}{\bibfnamefont{S.}~\bibnamefont{Jansen}},
  \bibinfo{author}{\bibfnamefont{M.~B.} \bibnamefont{Ruskai}},
  \bibnamefont{and} \bibinfo{author}{\bibfnamefont{R.}~\bibnamefont{Seiler}},
  \bibinfo{journal}{arXiv:quant-ph/0603175}  (\bibinfo{year}{2006}).

\bibitem[{\citenamefont{Schaller et~al.}(2006)\citenamefont{Schaller, Mostame,
  and Sch{\"u}tzhold}}]{Schaller}
\bibinfo{author}{\bibfnamefont{G.}~\bibnamefont{Schaller}},
  \bibinfo{author}{\bibfnamefont{S.}~\bibnamefont{Mostame}}, \bibnamefont{and}
  \bibinfo{author}{\bibfnamefont{R.}~\bibnamefont{Sch{\"u}tzhold}},
  \bibinfo{journal}{Physical Review A} \textbf{\bibinfo{volume}{73}}
  (\bibinfo{year}{2006}).

\bibitem[{\citenamefont{Joye}(2006)}]{Joye}
\bibinfo{author}{\bibfnamefont{A.}~\bibnamefont{Joye}},
  \bibinfo{journal}{arXiv:math-ph/0608059}  (\bibinfo{year}{2006}).

\bibitem[{\citenamefont{Aharonov et~al.}(2004)\citenamefont{Aharonov, van Dam,
  Kempe, Landau, Lloyd, and Regev}}]{Aharonov}
\bibinfo{author}{\bibfnamefont{D.}~\bibnamefont{Aharonov}},
  \bibinfo{author}{\bibfnamefont{W.}~\bibnamefont{van Dam}},
  \bibinfo{author}{\bibfnamefont{J.}~\bibnamefont{Kempe}},
  \bibinfo{author}{\bibfnamefont{Z.}~\bibnamefont{Landau}},
  \bibinfo{author}{\bibfnamefont{S.}~\bibnamefont{Lloyd}}, \bibnamefont{and}
  \bibinfo{author}{\bibfnamefont{O.}~\bibnamefont{Regev}},
  \bibinfo{journal}{FOCS}  (\bibinfo{year}{2004}),
  \bibinfo{note}{arXiv:quant-ph/0405098}.

\bibitem[{\citenamefont{Feynman}(1985)}]{Feynman}
\bibinfo{author}{\bibfnamefont{R.}~\bibnamefont{Feynman}},
  \bibinfo{journal}{Optics News} pp. \bibinfo{pages}{11--20}
  (\bibinfo{year}{1985}), \bibinfo{note}{reprinted in Foundations of Physics
  16(6) 507-531, 1986}.

\bibitem[{\citenamefont{Kitaev et~al.}(2002)\citenamefont{Kitaev, Shen, and
  Vyalyi}}]{Kitaev}
\bibinfo{author}{\bibfnamefont{A.~Y.} \bibnamefont{Kitaev}},
  \bibinfo{author}{\bibfnamefont{A.~H.} \bibnamefont{Shen}}, \bibnamefont{and}
  \bibinfo{author}{\bibfnamefont{M.~N.} \bibnamefont{Vyalyi}},
  \emph{\bibinfo{title}{Classical and Quantum Computation}},
  vol.~\bibinfo{volume}{47} of \emph{\bibinfo{series}{Graduate Studies in
  Mathematics}} (\bibinfo{publisher}{American Mathematical Society},
  \bibinfo{year}{2002}).

\bibitem[{\citenamefont{Kempe et~al.}(2004)\citenamefont{Kempe, Kitaev, and
  Regev}}]{Kempe}
\bibinfo{author}{\bibfnamefont{J.}~\bibnamefont{Kempe}},
  \bibinfo{author}{\bibfnamefont{A.}~\bibnamefont{Kitaev}}, \bibnamefont{and}
  \bibinfo{author}{\bibfnamefont{O.}~\bibnamefont{Regev}},
  \bibinfo{journal}{Proceedings of FSTTCS}  (\bibinfo{year}{2004}),
  \bibinfo{note}{arXiv:quant-ph/0406180}.

\bibitem[{\citenamefont{Childs et~al.}(2001)\citenamefont{Childs, Farhi, and
  Preskill}}]{Childs}
\bibinfo{author}{\bibfnamefont{A.}~\bibnamefont{Childs}},
  \bibinfo{author}{\bibfnamefont{E.}~\bibnamefont{Farhi}}, \bibnamefont{and}
  \bibinfo{author}{\bibfnamefont{J.}~\bibnamefont{Preskill}},
  \bibinfo{journal}{Physical Review A} \textbf{\bibinfo{volume}{65}}
  (\bibinfo{year}{2001}).

\bibitem[{\citenamefont{Sarandy and Lidar}(2005)}]{Sarandy}
\bibinfo{author}{\bibfnamefont{M.~S.} \bibnamefont{Sarandy}} \bibnamefont{and}
  \bibinfo{author}{\bibfnamefont{D.~A.} \bibnamefont{Lidar}},
  \bibinfo{journal}{Physical Review Letters} \textbf{\bibinfo{volume}{95}},
  \bibinfo{pages}{250503} (\bibinfo{year}{2005}),
  \bibinfo{note}{arXiv:quant-ph/0502014}.

\bibitem[{\citenamefont{{\AA}berg et~al.}(2005)\citenamefont{{\AA}berg, Kult,
  and Sj{\"o}qvist}}]{Aberg}
\bibinfo{author}{\bibfnamefont{J.}~\bibnamefont{{\AA}berg}},
  \bibinfo{author}{\bibfnamefont{D.}~\bibnamefont{Kult}}, \bibnamefont{and}
  \bibinfo{author}{\bibfnamefont{E.}~\bibnamefont{Sj{\"o}qvist}},
  \bibinfo{journal}{Physical Review A} \textbf{\bibinfo{volume}{72}},
  \bibinfo{pages}{042317} (\bibinfo{year}{2005}),
  \bibinfo{note}{arXiv:quant-ph/0507010}.

\bibitem[{\citenamefont{Roland and Cerf}(2005)}]{Roland}
\bibinfo{author}{\bibfnamefont{J.}~\bibnamefont{Roland}} \bibnamefont{and}
  \bibinfo{author}{\bibfnamefont{N.~J.} \bibnamefont{Cerf}},
  \bibinfo{journal}{Physical Review A} \textbf{\bibinfo{volume}{71}},
  \bibinfo{pages}{032330} (\bibinfo{year}{2005}),
  \bibinfo{note}{arXiv:quant-ph/0409127}.

\bibitem[{\citenamefont{Kaminsky and Lloyd}(2003)}]{Kaminsky}
\bibinfo{author}{\bibfnamefont{W.~M.} \bibnamefont{Kaminsky}} \bibnamefont{and}
  \bibinfo{author}{\bibfnamefont{S.}~\bibnamefont{Lloyd}}, in
  \emph{\bibinfo{booktitle}{Quantum Computing and Quantum Bits in Mesoscopic
  Systems}} (\bibinfo{publisher}{Kluwer Academic}, \bibinfo{year}{2003}),
  \bibinfo{note}{arXiv:quant-ph/0211152}.

\bibitem[{\citenamefont{Bacon et~al.}(2001)\citenamefont{Bacon, Brown, and
  Whaley}}]{Bacon}
\bibinfo{author}{\bibfnamefont{D.}~\bibnamefont{Bacon}},
  \bibinfo{author}{\bibfnamefont{K.~R.} \bibnamefont{Brown}}, \bibnamefont{and}
  \bibinfo{author}{\bibfnamefont{K.~B.} \bibnamefont{Whaley}},
  \bibinfo{journal}{Physical Review Letters} \textbf{\bibinfo{volume}{87}},
  \bibinfo{pages}{247902} (\bibinfo{year}{2001}),
  \bibinfo{note}{arXiv:quant-ph/0012018}.

\bibitem[{\citenamefont{Gottesman}(1997)}]{Gottesman}
\bibinfo{author}{\bibfnamefont{D.}~\bibnamefont{Gottesman}}, Ph.D. thesis,
  \bibinfo{school}{Caltech} (\bibinfo{year}{1997}),
  \bibinfo{note}{arXiv:quant-ph/9705052}.

\bibitem[{\citenamefont{Nielsen and Chuang}(2000)}]{Nielsen}
\bibinfo{author}{\bibfnamefont{M.~A.} \bibnamefont{Nielsen}} \bibnamefont{and}
  \bibinfo{author}{\bibfnamefont{I.~L.} \bibnamefont{Chuang}},
  \emph{\bibinfo{title}{Quantum Computation and Quantum Information}}
  (\bibinfo{publisher}{Cambridge University Press}, \bibinfo{year}{2000}).

\end{thebibliography}

\end{document}